\documentclass[12pt,a4j]{article}
\usepackage[dvipdfmx]{graphicx, color}
\usepackage{amsmath,enumerate, amsfonts, ulem, url}
\usepackage{natbib}
\newcommand{\bm}[1]{\mbox{\boldmath{$#1 $}}}

\newtheorem{prop}{Proposition}
\begin{document}
	\baselineskip 18pt
	\begin{center}
		{\LARGE\textbf{Truncated estimation for varying-coefficient functional linear model}}
	\end{center}
	\begin{center}
		{\large Hidetoshi Matsui}
	\end{center}
	
	\begin{center}
		\begin{minipage}{14cm}
			{
				\begin{center}
					{\it {\footnotesize 
							Faculty of Data Science, Shiga University \\
							1-1-1, Banba, Hikone, Shiga, 522-8522, Japan. \\
							hmatsui@biwako.shiga-u.ac.jp\\
					}}
					
				\end{center}
				%-----------------------------------------------
				\vspace{1mm} 
				{\small {\bf Abstract:}
					Varying-coefficient functional linear models consider the relationship between a response and a predictor, where the response depends not only the predictor but also an exogenous variable. 
					It then accounts for the relation of the predictors and the response varying with the exogenous variable.  
					We consider the method for truncated estimation for varying-coefficient models.  
					The proposed method can clarify to what time point the functional predictor relates to the response at any value of the exogenous variable by investigating the coefficient function.  
					To obtain the truncated model, we apply the penalized likelihood method with the sparsity inducing penalty.   
					Simulation studies are conducted to investigate the effectiveness of the proposed method. 
					We also report the application of the proposed method to the analysis of crop yield data to investigate when an environmental factor relates to the crop yield at any season.
				}
				
				\vspace{3mm}
				
			}
		\end{minipage}
	\end{center}
	
	%%%%%%%%%%%%%%%%%%%%%%%%%%%%%%%%%%%%%%%%%%%%%%%%%%%%%%%%%%%%%%%%%%%%%%%%%
	\section{Introduction}
	
	Functional data analysis is one of the most useful tools for analyzing longitudinal data, and comprehensive works for it have been developed in viewpoints of methodology and theory \cite{RaSi2005,HoKo2012,HsEu2015,KoRe2017}.  
	It also shows effectiveness in a various field of applications such as medicine, biology and meteorology \cite{RaSi2002,SoGoSa2013,Ullah2013}.  
	Functional regression analysis, a branch of functional data analysis, considers the relationship between predictors and responses where both or either of them are given as functions by using the functional regression models.  
	Enormous research for the functional regression models is reported in \cite{Mo2015,ReGoSh_etal2017}.  
	One of the key features of the functional linear model is that the coefficients in the model are given as functions rather than scalar, and then we can investigate how and when the predictor relates to the response.  

    Let $X(s)$ be a functional predictor on $s\in\mathcal S\subset\mathbb R$ and $Y$ a scalar response. 
    Then the traditional scalar-on-function functional linear model is given by 
    \begin{align}
	    Y = \int_{\mathcal S} X(s)\beta(s)ds + \varepsilon,
	    \label{eq:FLM}
	\end{align}
	where $\beta(s)$ is a coefficient function and $\varepsilon$ is a random noise.  
	By estimating $\beta(s)$, we can investigate the relation between $X$ and $Y$ at arbitrary time point $s$.  
	In the process of estimating the functional linear model (\ref{eq:FLM}), the assumption that the coefficient functions are represented by basis expansions if often used \citep{CaCrKn_etal2007,ArKoKa_etal2009a}.  
	\cite{JaWaZh2009} proposed imposing sparsity-inducing penalties \citep{HaTiWa2015} on the coefficients of the basis expansion, and then proposed a method for the selection of the domain of the functional predictor, that is, their method shrinks $\beta(s)$ toward exactly zero at some intervals on $\mathcal S$ in the model (\ref{eq:FLM}).  
	Then we can investigate how and when the functional predictor relates to the response.  
	\cite{LePa2012, ZhWaWa2013, LiCaWa_etal2017, PiSeVi2019} also approached this problem.  
	\cite{KoBoWu2015} proposed the method for the domain selection for the varying-coefficient models.  
	On the other hand, \cite{HaHo2016} proposed the method for estimating the functional linear model that clarifies to what time point the functional predictor relates to the response, by truncating values of the coefficient function after a certain time point toward exactly zeros.  
	They called this model the truncated functional linear model.
	The method of \cite{HaHo2016} does not use the sparse regularization, while more recently, \cite{GuLiCa_etal2020} proposed estimating the truncated functional linear model by applying the sparse regularization \citep{HuMaXi_etal2009}.  
    Note that the purpose of the application of sparse regularization in this work is different from variable selection of functional predictors, which investigates whether each variable itself is relevant to the response \citep{MaKo2011,GeMaSt2013,Ma2021}.  
    
    We consider the method for a truncated estimation for varying-coefficient functional linear models, which we call VCFLM.
    The VCFLM proposed by \cite{CaSa2008,WuFaMu2010} represents the relationship between a functional predictor and a scalar response, where the response also depends on another variable, called an exogenous variable.
    That is, the VCFLM is an extension of the functional linear model and the varying-coefficient model.
    Then we can clarify to what time point the functional predictor relates to the response at any value of the exogenous variable by investigating the coefficient function of the VCFLM.   
    The motivation comes from the analysis of data for crop yield of tomatoes.  
    A plant of multi-stage tomatoes grows for almost one year, and fruits are harvested every day for a long term of the year.  
    It is said that the amount of the crop yield at certain day depends on environmental factors such as temperature, solar radiation at the farmland for a certain period of days before that day.  
    However, the length of the period has not quantitatively been investigated. 
    Furthermore, it is also said that this length differs for season; the period that the temperature relates to the crop yield differs in summer and winter.  
    Therefore, we express the relationship between the crop yield and the environmental factor by the VCFLM, treating the calendar time (day of the year) as the exogenous variable.  
    In addition, we apply the truncated estimation for the VCFLM to investigate how many days of the environmental factor relates to the crop yield at any season. 

	In this article we consider estimating a truncated varying-coefficient functional linear model.  
    We assume that the VCFLM is expressed by basis expansions, and then unknown parameters are estimated by the penalized likelihood method with the nested group bridge penalty using the idea of \cite{HuMaXi_etal2009} to obtain a truncated VCFLM.  
    We also show that the maximizer of the penalized likelihood with the nested group bridge penalty coincides with that with the traditional lasso penalty and that the estimation algorithm has similar form as the lasso.  
    To select tuning parameters included in the penalized likelihood method, we use a model selection criterion derived by using the idea of \cite{ZhLiTs2010}.  
    To investigate the effectiveness of the proposed method, we conduct some simulation studies, and then we apply the proposed method to the analysis of crop yield data. 
	
	This article is organized as follows. Section 2 reviews the details of the VCFLM, and then Section 3 describes the method for estimating the VCFLM via the sparse regularization to obtain the truncated model.  
    Simulation studies are conducted in Section 4, and then we report the result of real data analysis in Section 5.  
	Finally, we conclude the main points and give future works in Section 6.  
	%%%%%%%%%%%%%%%%%%%%%%%%%%%%%%%%%%%%%%%%%%%%%%%%%
	\section{Varying-coefficient functional linear model}
	Suppose we have $n$ sets of observations $\{x_{i}(s), y_i, t_i;~ i=1,\ldots, n, ~s\in\mathcal S\subset \mathbb R, ~t_i\in\mathcal T\subset\mathbb R \}$, where $x_{i}(s)$ is a functional predictor and $y_i=y_i(t_i)$ is a scalar response that depends on an exogenous variable $t_i$.  
	The functional predictor and the scalar response are centered so that the mean value of the response and the mean function of the predictor become zero, respectively.  
	Then we model the relationship between the predictor and the  response by the varying-coefficient functional linear model \citep[VCFLM,][]{CaSa2008, MuYa2008}: 
	\begin{align}
		y_i(t_i) =\int_{\mathcal S} x_{i}(s)\beta(s, t_i)ds + \varepsilon_i, 
		\label{eq:VCFLM}
	\end{align}
	where $\beta(s, t)$ is a bivariate coefficient surface with respect to $s$ and $t$, and $\varepsilon_i$ is an independently and identically normally distributed error with mean 0 and variance $\sigma^2$.
	%, where $\bm{\varepsilon} = (\varepsilon_1,\ldots, \varepsilon_n)^T$ has mean zero vector and variance-covariance matrix $\Sigma.$  
	The coefficient function $\beta(s,t)$ provides us insights about how the functional predictor $x_{i}(s)$ affects the response $y_i$ at arbitrary exogenous variable $t_i$.  
	
	Since observed data for the functional predictor $x_i(s)$ are repeatedly measured at discrete time points in nature, we need to transform them into functions.  
	We assume that the longitudinal data corresponding to the predictor are transformed into functions by basis expansions as follows.
	\begin{align}
		x_{i}(s) = \sum_{k=1}^{m^{(1)}} w_{ik}\phi_{k}(s) = \bm w_{i}^T\bm{\phi}(s), 
		\label{eq:x}
	\end{align}
	where $m^{(1)}$ is the number of basis functions, $\bm\phi(s) = (\phi_{1}(s),\ldots,\phi_{m^{(1)}}(s))^T$ is a vector of basis functions and $\bm w_{i} = (w_{i1},\ldots, w_{im^{(1)}})^T$ is a vector of coefficient for each subject $i$.  
	Here $\bm\phi(s)$ must be a set of sequential and support compact basis functions, and here we use the $B$-splines \citep{de2001}.  
	The $\bm w_{i}$ is obtained by the smoothing method via the penalized likelihood method in advance, and therefore they are supposed to be known here.  
	%%%%%%%%%%%%%%%%%%%%%%%%%%%%%%%%%%%%%%%%%%%%%%%%%%%%%%%%%%%%%%%%%%%%%%%%%
	
\section{Estimation}
To estimate the coefficient function $\beta(s,t)$ in the VCFLM (\ref{eq:VCFLM}), we assume that and $\beta(s,t)$ are expressed by basis expansions such as
\begin{align}
	\beta(s,t) &= \sum_{k=1}^{m^{(1)}}\sum_{l=1}^{m^{(2)}}\phi_{k}(s) b_{kl} \psi_{l}(t) 
	= \bm{\phi}(s)^T B\bm{\psi}(t),
	\label{eq:beta}
\end{align}
where $\bm{\psi}(t) = (\psi_{1}(t),\ldots, \psi_{m^{(2)}}(t))^T$ is a vector of $m^{(2)}$ basis functions, and $B = (b_{kl})_{kl}$ is a matrix of unknown parameters.
By the above assumptions of basis expansions (\ref{eq:x}) and (\ref{eq:beta}), the VCFLM is expressed by
\begin{align}
	y_i 
	&= \bm w_{i}^T\Phi B\bm{\psi}(t_i) + \varepsilon_i \nonumber \\
	&= \left\{\bm{\psi}(t_i)^T\otimes \bm w_{i}^T\Phi\right\}{\rm vec}B + \varepsilon_i \nonumber \\
	&= \bm z_i^T\bm{b} + \varepsilon_i,
	\label{eq:VCFLM2}
\end{align}
where $\Phi = \int_{\mathcal T}\bm{\phi}(s)\bm{\phi}(s)^Tds$,  $\otimes$ denotes a Kronecker product of the matrix, ${\rm vec}$ is an operator that align a matrix to a vector in the column-wise.   
Furthermore, $\bm z_i = (\bm{\psi}(t_i)^T\otimes \bm w_{i}^T\Phi)^T$ is a known vector and $\bm b = {\rm vec}B$ is an unknown vector.  
Therefore, the problem of estimating the VCFLM (\ref{eq:VCFLM}) becomes that of estimating $\bm{b}$ in (\ref{eq:VCFLM2}) and we can estimate it by traditional estimation strategies such as least squares or maximum likelihood method.  
Under the above assumptions, we have a statistical model for the response $y_i$ as follows:
%vector $\bm y = (y_1,\ldots, y_n)^T$ as follows:  
\begin{align}
%		f(\bm y; \bm{\theta}, \Sigma) = \frac{1}{(2\pi)^{n/2}|\Sigma|^{1/2}}
%		\exp\left\{-\frac{1}{2}(\bm y - Z\bm{\theta})^T\Sigma^{-1}(\bm y - Z\bm{\theta})\right\}, 
	f(y_i(t_i); \bm{\theta}) = \frac{1}{\sqrt{2\pi\sigma^2}}
	\exp\left\{-\frac{1}{2\sigma^2}(y_i(t_i) - \bm z_i^T\bm{\theta})^2\right\}, 
	\label{eq:statmodel}
\end{align}
%	where $Z = (\bm z_1,\ldots, \bm z_n)^T.$  
where $\bm{\theta}$ $=$ $(\bm b^T, \sigma^2)^T$ is a vector of unknown parameters.

We estimate the unknown parameter $\bm b$ by the penalized maximum likelihood method with the nested group bridge and smoothness penalties.  
Let $\bm b_l = (b_{1l},\ldots, b_{m^{(1)} l})^T$ denotes the column-wise vectors of $B$, and let $A_{k} = \{k, k+1,\ldots, m_1+d\}$ and $\bm b_{A_k, l} = (b_{kl}, b_{k+1,l},\ldots, b_{m^{(1)}+d, l})^T$.
Then we estimate the coefficient parameters by maximizing the following penalized log-likelihood function:
\begin{align}
	\ell_{\kappa,\lambda}(\bm{\theta}) = \ell(\bm{\theta}) - 
	n\kappa\sum_{l=1}^{m^{(2)}} \|\bm b_l^T\bm{\phi}^{(2)}\|^2 - 
%		n\kappa_2\sum_{k=1}^{m^{(1)}} \|\bm b_{(k)}^T\bm{\psi}^{(2)}(t)\|^2 - 
	n\lambda\sum_{l=1}^{m^{(2)}}\sum_{k=1}^{m^{(1)}}c_{kl}\|\bm b_{A_k,l}\|_1^{\gamma},
	\label{eq:NGB}
\end{align}
where $\ell(\bm{\theta}) = \sum_{i=1}^{n}\log f(y_i(t_i); \bm{\theta})$ is a log-likelihood function, and $\bm{\phi}^{(2)}(s)$ is a vector of second derivatives of basis functions $\bm{\phi}(s)$.  
%	●第２項はs方向にしか制約かけてない　t方向への正則化は立式が複雑になるので、基底の個数で調整. 
The $\kappa>0$ is a tuning parameter that controls the smoothness of the coefficient function $\beta(s,t)$ with respect to the $s$ direction, and $\lambda>0$ is a tuning parameter that controls the sparseness of the parameter $b_{kl}$.
Furthermore, $\gamma\in [0,1]$ is a tuning parameter for the group bridge penalty, and $c_{kl}$ is an adaptive weight that is given by $c_{kl} = |A_{kl}|^{1-\gamma} / \|{\check{\bm b}}_{A_{k},l}\|_1^\gamma$, where $\check{\bm b}$ is given by a ridge estimator.  
Then we can shrink some of the last elements of the vector $\bm b_l$ toward exactly zeros for each $l$.  
That is, the first, second and third terms in (\ref{eq:NGB}) respectively represent the goodness of fit of the model for the data, the smoothness penalty for the $s$ direction of $\beta(s,t)$, and the penalty which leads to the truncated estimation.
Note that this method gives a truncated estimation for only the $s$ direction of the coefficient function $\beta(s,t)$, not for the $t$ direction.  

Since the penalized log-likelihood function (\ref{eq:NGB}) contains the nonconcave group bridge penalty, it is difficult to construct an algorithm for finding a maximizer of (\ref{eq:NGB}).  
On the other hand, \cite{BrHu2009} showed that the maximizer of the log-likelihood function with the group bridge penalty coincides with that with the lasso-type penalty.  
Using the result of \cite{BrHu2009} straightforwardly, we have the following result.
\begin{prop}
	Let $\tau>0$ is a tuning parameter that satisfies $\lambda = \tau^{1-\gamma}\gamma^{-\gamma}(1-\gamma)^{\gamma -1}$. 
	Then the maximizer of the penalized log-likelihood function with the group bridge penalty (\ref{eq:NGB}) is the same as that of the following penalized log-likelihood function.
	\begin{align}
		\ell_{\kappa,\tau}(\bm{\theta}, \bm\eta) = 
		-\frac{1}{\sigma^2}\|{\bm y} - Z\bm b\|^2 
		- n\kappa\sum_{l=1}^{m^{(2)}} \|\bm b_l^T\bm{\phi}^{(2)}(s)\|^2
		- n\sum_{l=1}^{m^{(2)}}\sum_{k=1}^{m^{(1)}} c_{kl}^{1/\gamma}\eta_{kl}^{1-1/\gamma} \|\tilde{\bm b}_{A_{k},l}\|_1
		- n\tau \sum_{l=1}^{m^{(2)}}\sum_{k=1}^{m^{(1)}}\eta_{kl},
		\label{eq:NGB2}
	\end{align}
	where $\bm{\eta} = (\eta_{11}, \eta_{21},\ldots, \eta_{m^{(1)}m^{(2)}})^T$.  
	Note that the terms that do not depend on the parameter $\bm{\theta}$ are omitted in $\ell_{\kappa,\tau}(\bm{\theta}, \bm\eta)$.  
\end{prop}
The proof is given in Appendix.  
Let $V = \int \bm{\phi}^{(2)}(s)\bm{\phi}^{(2)}(s)^T ds$, then $\sum_{l=1}^{m^{(2)}}\|\bm b_l^T\bm{\phi}^{(2)}\|^2 = \bm b^T(I_{m^{(1)}}\otimes V)\bm b$.  
Since $V$ is a positive semi-definite matrix, we can obtain the matrix $W$ such that $I_{m^{(1)}}\otimes V = WW^T$ by the spectral decomposition.  
Furthermore, we denote
\begin{align*}
	\tilde{\bm y} = 
	\begin{pmatrix}
		\bm y \\ \bm 0_{m^{(1)}m^{(2)}}
	\end{pmatrix}, ~~
	U_*  = 
	\begin{pmatrix}
		\bm Z \\ \sqrt{n\sigma^2\kappa}W^T
	\end{pmatrix},~~
	g_{kl} = \sum_{j=1}^{k} c_{jl}^{1/\gamma}\eta_{jl}^{1-1/\gamma}.
\end{align*}
Then (\ref{eq:NGB2}) can be expressed by
\begin{align*}
	\ell_{\kappa,\tau}(\bm{\theta}, \bm{\eta}) = 
	-\frac{1}{\sigma^2}\|\tilde{\bm y} - U_*\bm b\|^2
	- \sum_{l=1}^{m^{(2)}}\sum_{k=1}^{m^{(1)}}g_{kl}|b_{kl}| 
	- \tau\sum_{l=1}^{m^{(2)}}\sum_{k=1}^{m^{(1)}}\eta_{kl},
\end{align*}
Furthermore, denoting $G = {\rm diag}\{(ng_{11})^{-1}, (ng_{21})^{-1}, \ldots, (ng_{m^{(1)}m^{(2)}})^{-1}\}$, $\tilde U = U_*G$ and $\tilde{\bm b} = G^{-1}\bm b$, then $\ell_{\kappa,\tau}(\bm{\theta}, \bm{\eta})$ is re-expressed by
\begin{align}
\ell_{\kappa,\tau}(\tilde{\bm{\theta}}, \bm{\eta}) = 
-\frac{1}{\sigma^2}\|\tilde{\bm y} - \tilde U\tilde{\bm b}\|^2
- \sum_{l=1}^{m^{(2)}}|\tilde b_{kl}| 
- \tau\sum_{l=1}^{m^{(2)}}\sum_{k=1}^{m^{(1)}}\eta_{kl},
\label{eq:olasso}
\end{align}
where $\tilde{\bm{\theta}} = (\tilde{\bm b}^T, \sigma^2)^T$.  
That is, the problem of estimating the parameter vector $\tilde{\bm{\theta}}$ by maximizing the penalized log-likelihood function (\ref{eq:NGB}) becomes that by maximizing (\ref{eq:olasso}).  
Although $\bm{\eta}$ is not used for estimating the VCFLM, the unknown parameters $\tilde{\bm{\theta}}$ and $\bm{\eta}$ are estimated by iteratively updating each other. 
The coefficient parameter $\tilde{\bm{b}}$ included in $\tilde{\bm{\theta}}$ is estimated by applying the ordinary lasso estimation algorithm, here we applied the coordinate descent algorithm \citep{FrHaTi2010a}.
On the other hand, $\tilde{\bm b}$ is updated in an analytical expression.  

The estimator is obtained by the following algorithm:
\begin{enumerate}
	\item Set initial values for $\bm\theta$ and denote it as $\bm b^{(0)} = (\bm b^{(0)}, \sigma^{(0)2})^T$.  
	Set $h=0.$   
	\item Calculate the $h$-th update values of $\eta_{kl}$, $g_{kl}$ $(j=1,\ldots, m^{(1)}, k=1,\ldots, m^{(2)})$, and $\tilde U$, respectively, as 
	\begin{align*}
		\eta_{kl}^{(h)} &= c_{kl}\left(\frac{1-\gamma}{\tau\gamma}\right)^\gamma
		\|\bm b_{A_k, l}^{(h)}\|_1^\gamma, \\
		g_{kl}^{(h)} &= 
		\sum_{j=1}^{k}\left(\eta_{jl}^{(h)}\right)^{1-1/\gamma}c_{kl}^{{1/\gamma}}, \\
		\tilde U^{(h)} &= U_*G^{(h)}, ~~
		G^{(h)} = {\rm diag}\left\{1/g_{11}^{(h)}, 1/g_{12}^{(h)}, \ldots, 1/g_{m^{(1)}m^{(2)}}^{(h)}\right\}. 
	\end{align*}
	\item Derive an updated value $\tilde{\bm b}^{(h+1)}$ by solving the ordinal lasso problem
	\begin{align*}
		\min_{\tilde b} \frac{1}{\sigma^{(h)2}}\|\tilde{\bm y} - \tilde U^{(h)}\tilde{\bm b}\|_2^2 + \sum_{l=1}^{m^{(2)}}\sum_{k=1}^{m^{(1)}}|\tilde b_{kl}|,
	\end{align*}
	and then update ${\bm b}^{(h+1)}$ by ${\bm b}^{(h+1)} = G^{(h)}\tilde{\bm b}^{(h)}$.  
	\item Update the variance parameter $\sigma^2$ by
	\begin{align*}
		\sigma^{(h+1)2} = \frac{1}{n}\|\tilde{\bm y} - U_*{\bm b}^{(h+1)}\|^2.
	\end{align*}
	\item Iterate steps 2. to 4. until convergence.  
\end{enumerate}
	
The statistical model (\ref{eq:statmodel}) estimated by the penalized likelihood method with the nested group bridge depends on tuning parameters such as the number $m^{(1)}$ and $m^{(2)}$ of basis functions, $\kappa$, $\lambda$ and $\gamma$.  
To select these values, we apply a BIC-type model selection criterion \citep{Sc1978}.  
Using the result of, \cite{ZhLiTs2010}, the BIC for evaluating the truncated VCFLM estimated by the nested group bridge penalty is given by
\begin{align*}
	{\rm BIC} = -2\ell(\hat{\bm{\theta}}) + 2\widehat{edf},
\end{align*}
where $\widehat{edf}$ is an effective degrees of freedom and is given by 
\begin{align*}
	\widehat{edf} = {\rm tr}\left\{Z^T\left(Z^TZ + n\kappa W_{\mathcal A}W_{\mathcal A}^T\right)^{-1}Z\right\},  
\end{align*}
here $W_{\mathcal A}$ is a part of the matrix $W$ for active set $\mathcal A$.  
We select the tuning parameters that minimize the BIC, and then decide the corresponding model as an optimal one.  

%%%%%%%%%%%%%%%%%%%%%%%%%%%%%%%%%%%%%%%%%%%%%%%%%%%%%%%%%%%%%%%%%%%%%%%%%%%%%%%%%%
\section{Simulation study}
In this section we provide some results on Monte Carlo simulations to investigate the effectiveness of the proposed method.  
We numerically generate the dataset for a time-course predictor, a scalar response, and an exogenous variable, and then applied the proposed and existing methods to compare the performance of them.  

First, we generated the data for a functional predictor.  
In real applications, the data for the functional predictor are observed at discrete time points in nature rather than continuously, and therefore the longitudinal data are generated at discrete time points with additional Gaussian noises.  
That is, an observation for the functional predictor for the $i$-th subject at $\alpha$-th time point $s_\alpha\in\mathcal S=[0,1]$ $(\alpha = 1,\ldots, N)$ is given by
\begin{align*}
	x_{i\alpha} = g_{i}(s_\alpha) + e_i,~~  
	g_{i}(s) = \bm w_{i}^T\bm{\varphi}(s), 
\end{align*}
where $\bm w_{i}$ is a vector of normally distributed random coefficients, \bm{\varphi}(s) is a vector of $B$-spline basis functions, and $e_i$ is an i.i.d normally distributed random noise with mean zero and standard deviation $\tau = 0.1R_{ij}^X$ with $R_{i}^X = \max_\alpha g_{i}(s_\alpha) - \min_\alpha g_{i}(s_\alpha)$.  
The number of time points is set to be $N=21$ for all subject.  
We then apply the smoothing method to $x_{i1}, \ldots, x_{iN}$ using $B$-spline basis for all $i=1,\ldots, n$, and then obtain functional data $\{x_i(s); i=1\ldots, n\}.$
For more details about the smoothing method, see \cite{GrSi1994,Ko2014}.  
Here we used an R package {\tt fda} for transforming the longitudinal data into functional data.  

\begin{figure}[t]
	\begin{center}
		\includegraphics[width=0.48\hsize]{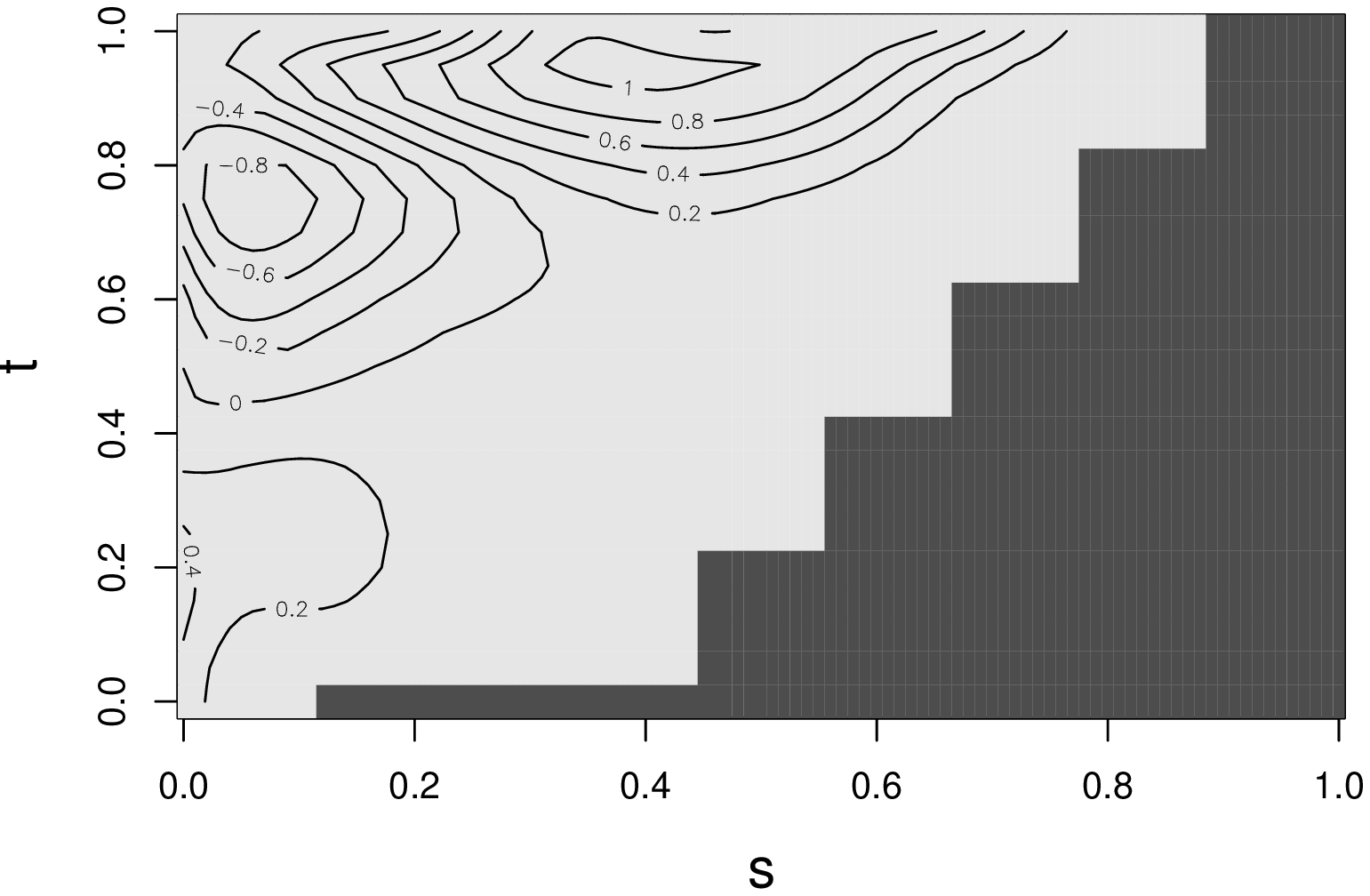}
		\includegraphics[width=0.48\hsize]{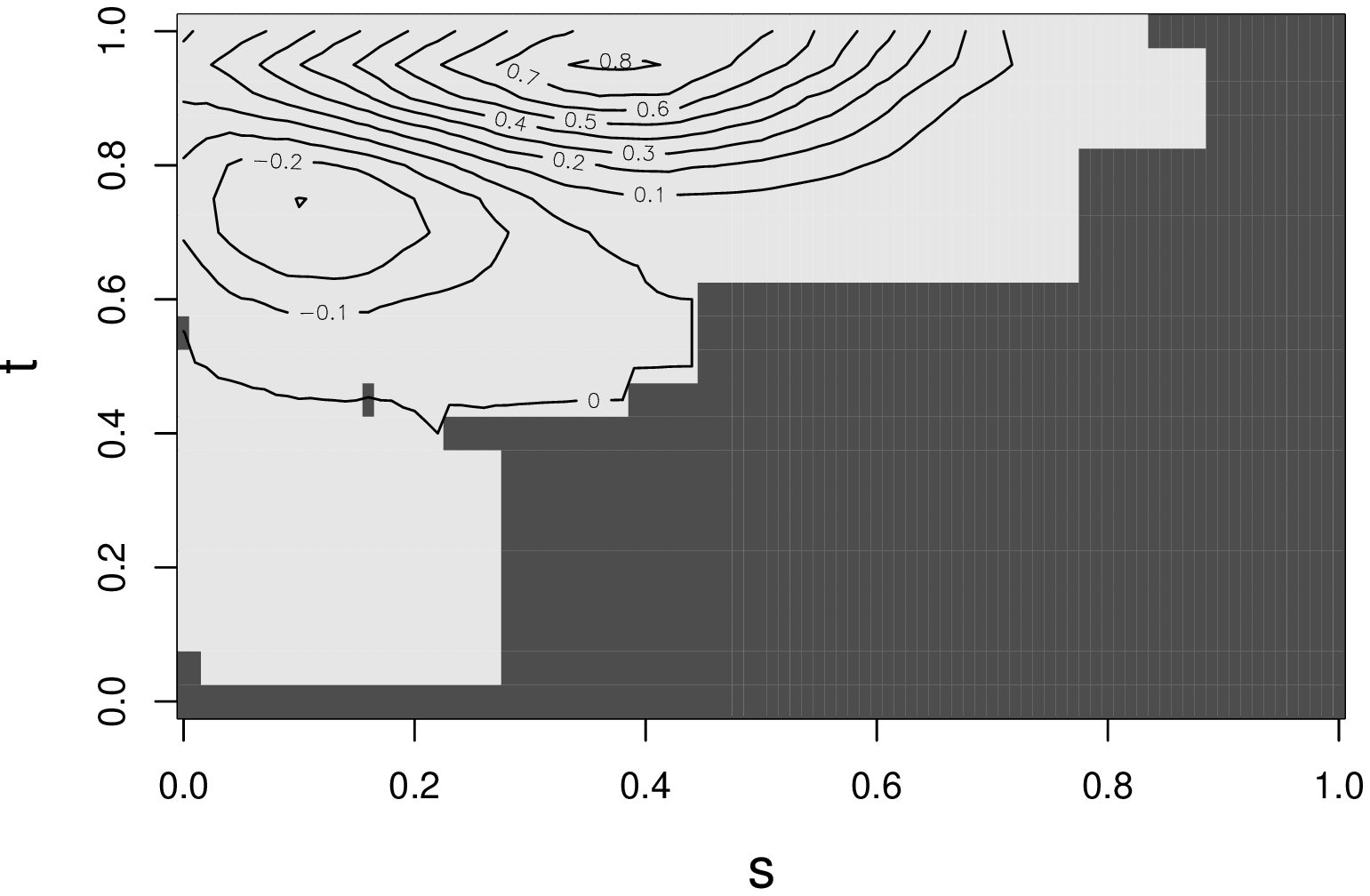} \\
		\includegraphics[width=0.48\hsize]{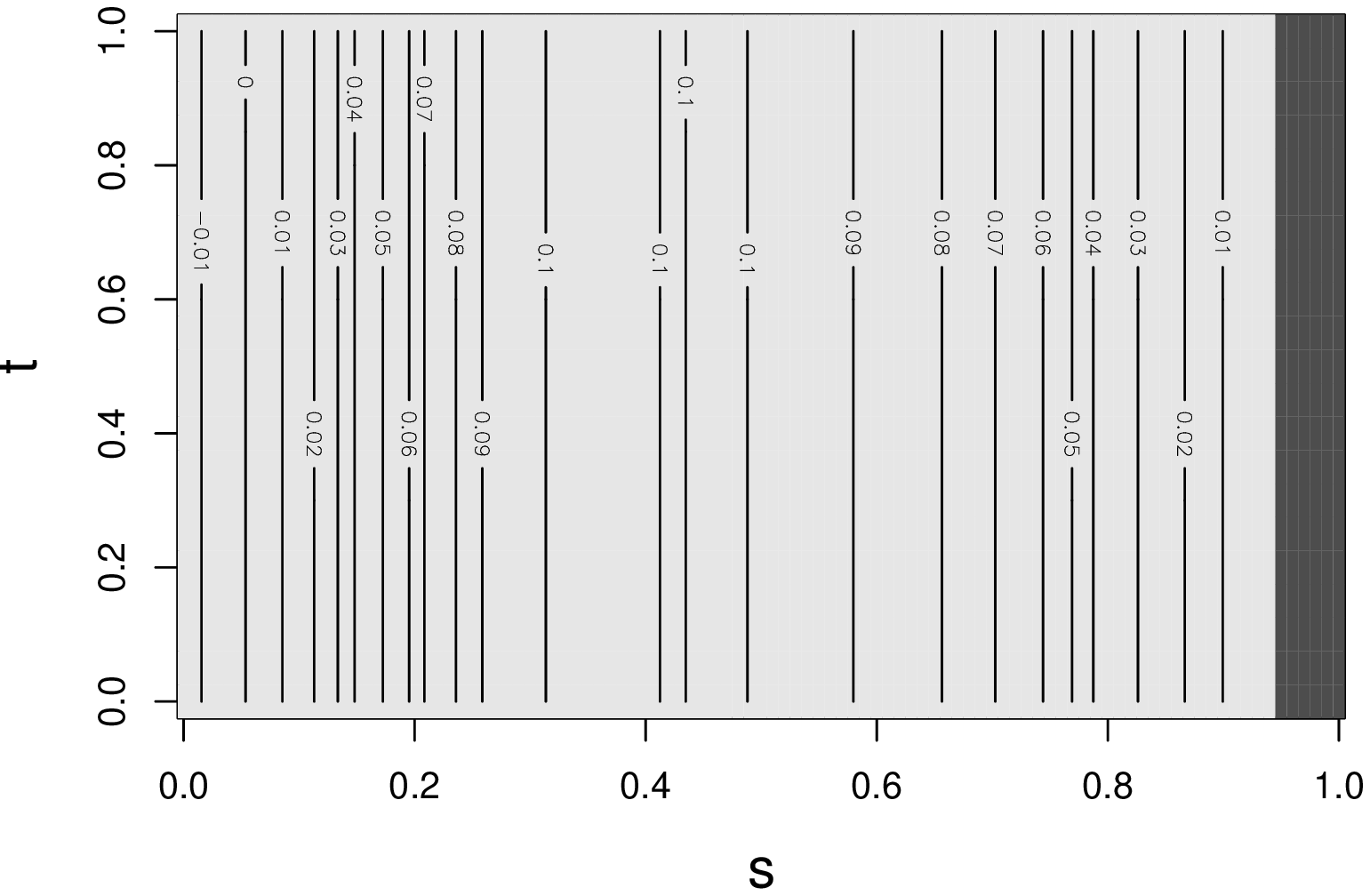}
		\includegraphics[width=0.48\hsize]{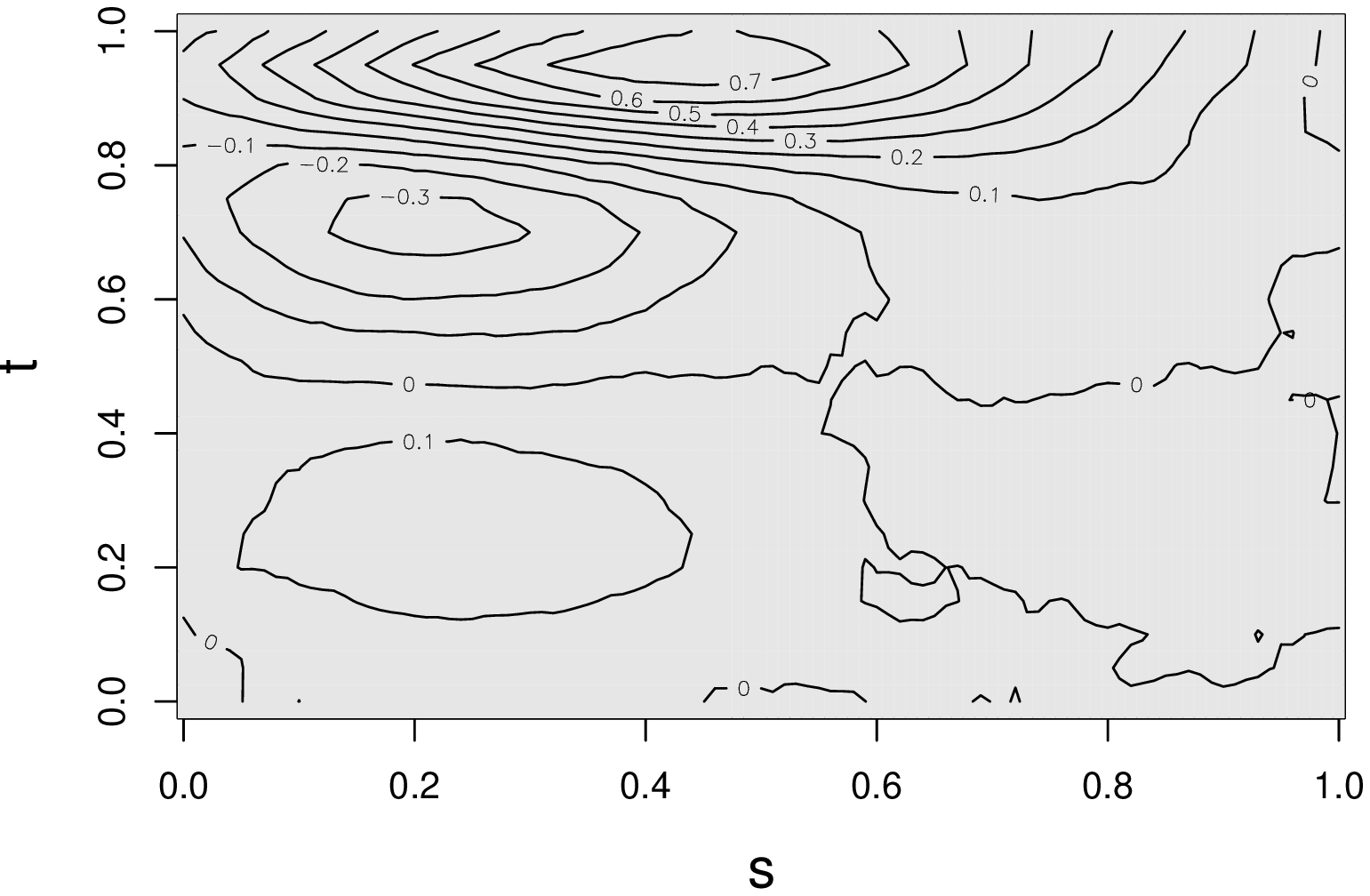}
	\end{center}
	\caption{True and estimated coefficient surfaces in simulation study.  Top left: True surface. Top right: TVCFLM.  Bottom left: TFLM, Bottom right: VCFLM.  Dark gray represents areas where the coefficient is estimated to be 0.} 
	\label{fig:beta_sim}
\end{figure}

The coefficient function $\beta(s,t)$ is supposed to be expressed by basis expansions as (\ref{eq:beta}), where the elements of the matrix $B$ are generated from normally distributed random values.  
Here, for the $s$ direction, all elements of $B$ after a certain time point are set to be 0 for each $t$, so that the truncated model is obtained.  
A contour plot of $\beta(s,t)$ is shown in the top left of Figure \ref{fig:beta_sim}.  
For simplicity, we fixed the number of basis functions as $m^{(1)}=21, m^{(2)}=8$ and $\gamma =0.5$, and we selected the remaining tuning parameters $\kappa, \tau$ by BIC.
Then we generated the response by the following VCFLM:
\begin{align*}
	y_{i}(t_i) = f(x_{i}) + \varepsilon_{i},~~
	f(x_{i}) = \int_{\mathcal S} x_{i}(s) \beta\left(s, t_{i}\right)ds, 
	%	\label{eq:VCFLM}
\end{align*}
where $t_i\in [0,1]$ is the exogenous variable.  
We assume in this simulation that the errors $\varepsilon_1, \ldots, \varepsilon_n$ in the VCFLM has mean zero and standard deviation $\sigma = rR^Y$, where $s$ indicates the noise level and here we set $r=0.1, 0.3$, and $R^Y = \max_i f(x_{i}) - \min_i f(x_{i})$.  
The purpose of this simulation is to evaluate the accuracy of the predicted response and the estimated coefficient of the proposed method, and in addition, whether the proposed method correctly shrinks a part of the domain of the coefficient toward exactly zero.  

For the generated data set, we apply the proposed truncated VCFLM (TVCFLM) to the generated data.  
We also compared the performance of the proposed method with the existing method; truncated functional linear model (TFLM) by  \cite{GuLiCa_etal2020} and VCFLM with the ridge-type penalized likelihood estimation.  
Here we used R package {\tt ngr} for implementing TFLM. 
In addition, since TFLM does not depend on the exogenous variable $t$, the estimated coefficient function is $\widehat\beta(s,t) \equiv \widehat\beta(s)$.
We then calculated the root mean squared errors (RMSE) for the response and coefficient surface, respectively given by
\begin{align*}
	{\rm RMSE}_y &= \left[\frac{1}{n}\sum_{i=1}^{n}\left\{
	f(x_{i}) - \widehat f(x_{i})
	\right\}^2\right]^{1/2}, \\
	{\rm RMSE}_\beta &= \left[\frac{1}{NT}\sum_{\alpha=1}^{N}\sum_{j=1}^{T}\left\{
	\beta(s_\alpha, t_{(j)}) - \widehat\beta(s_\alpha, t_{(j)}) 
	\right\}^2\right]^{1/2},  
\end{align*}
where $\widehat f$ and $\widehat\beta$ are estimated functions of $f$ and $\beta$, respectively, and $t_{(1)}, \ldots, t_{(T)}$ are equally spaced time points of $t$.  
We iterated this process 100 times, and then calculated the averaged values of 100 RMSEs.  
\begin{table}[t]
\centering
\caption{RMSE$_y$s $(\times 10^2)$ for simulation studies.  }
\label{tab:sim_y}
\begin{tabular}{lccccc}
   \hline
   & TVCFLM & TFLM & VCFLM \\ 
  \hline
  $r = 0.1$ \\
 $n=100$ & 7.35 (1.76) & 14.71 (1.95) & 7.64 (1.60) \\ 
 $n=200$ & 6.37 (1.25) & 14.98 (1.34) & 6.66 (0.99)\\ 
 $n=400$ & 5.65 (0.86) & 15.04 (0.82) & 5.91 (0.65)\\ 
  \hline
    $r = 0.3$ \\
 $n=100$    & 14.66 (4.04) & 15.18 (2.31) & 19.38 (4.79) \\ 
 $n=200$   & 12.70 (2.29) & 15.81 (1.76) & 17.06 (3.18)\\ 
 $n=400$   & 11.29 (1.75) & 15.90 (1.06) & 14.70 (2.52) \\
    \hline
\end{tabular}
\caption{RMSE$_\beta$s $(\times 10)$ for simulation studies.}
\label{tab:sim_beta}
\begin{tabular}{lcccc}
   \hline
   & TVCFLM & TFLM & VCFLM \\ 
  \hline
  $s = 0.1$ \\
 $n=100$    & 2.23 (0.55) & 2.89 (0.28) & 2.44 (0.41) \\ 
 $n=200$   & 1.80 (0.33) & 2.83 (0.18) & 2.08 (0.24)\\ 
 $n=400$   & 1.61 (0.22) & 2.76 (0.07) & 1.93 (0.17) \\
    \hline
  $s = 0.3$ \\
 $n=100$    & 3.23 (1.16) & 3.12 (0.41) & 4.80 (1.15) \\ 
 $n=200$   & 2.63 (0.42) & 3.00 (0.34) & 3.80 (0.70)\\ 
 $n=400$   & 2.39 (0.38) & 2.90 (1.87) & 3.22 (0.51) \\
    \hline
\end{tabular}
\end{table}

Tables \ref{tab:sim_y} and \ref{tab:sim_beta} shows averages (and standard deviations in brackets) of 100 RMSE$_y$s and RMSE$_\beta$s.   
Compared to the existing methods, the proposed method gives smaller RMSEs for almost all situations in viewpoints of both the response and the coefficient function.  
Since the TVCFLM and VCFLM use the same model, the difference of RMSEs between them is small when the noise level $r$ is small, and the VCFLM gives more stable estimates, since the number of parameters is smaller than TVCFLM.   
However, the RMSEs for VCFLM becomes worse when the noise is large.  
The TFLM does not consider the exogenous variable, and therefore the RMSEs are larger than other methods, but they are not much worse even when the noise level is large.  
In addition, the TFLM also gives smaller standard deviations. 

Figure \ref{fig:beta_sim} shows true coefficient function $\beta(s,t)$ and pointwise median for 100 estimated coefficient functions for three methods for the case $n=200$ and $r=0.1$.  
Note that we used the pointwise median rather than mean because we want to investigate whether the proposed method appropriately shrinks the coefficient surface at a subset of the domain toward exactly zero.  
This figure shows that the proposed method appropriately captures the trend of the true surface and shrinks the part of the surface towards exactly zero, which gives truncated model.  
The TFLM also gives a truncated model, but the domain of zero coefficient is very narrow compared to the true coefficient.  
Furthermore, in TFLM, the coefficient function in the t direction is fixed.  
Although the estimated coefficient by the VCFLM shows similar trend as the true one, it fails to obtain truncated model.  
%%%%%%%%%%%%%%%%%%%%%%%%%%%%%%%%%%%%%%%%%%%%%%%%%%%%%%%%%%%%%%%%%%%%%%%%%%%%%%%%%%
\section{Real data example}
We report the result of the analysis of the crop yield data for multi-stage tomatoes cultivated in a greenhouse on a farm in Kobe, Japan.  
The aim of this analysis is to investigate how many days before maturing the tomatoes the temperature relates to the crop yield, through the proposed method.  

\begin{figure}[t]
	\begin{center}
		\includegraphics[width=0.48\hsize]{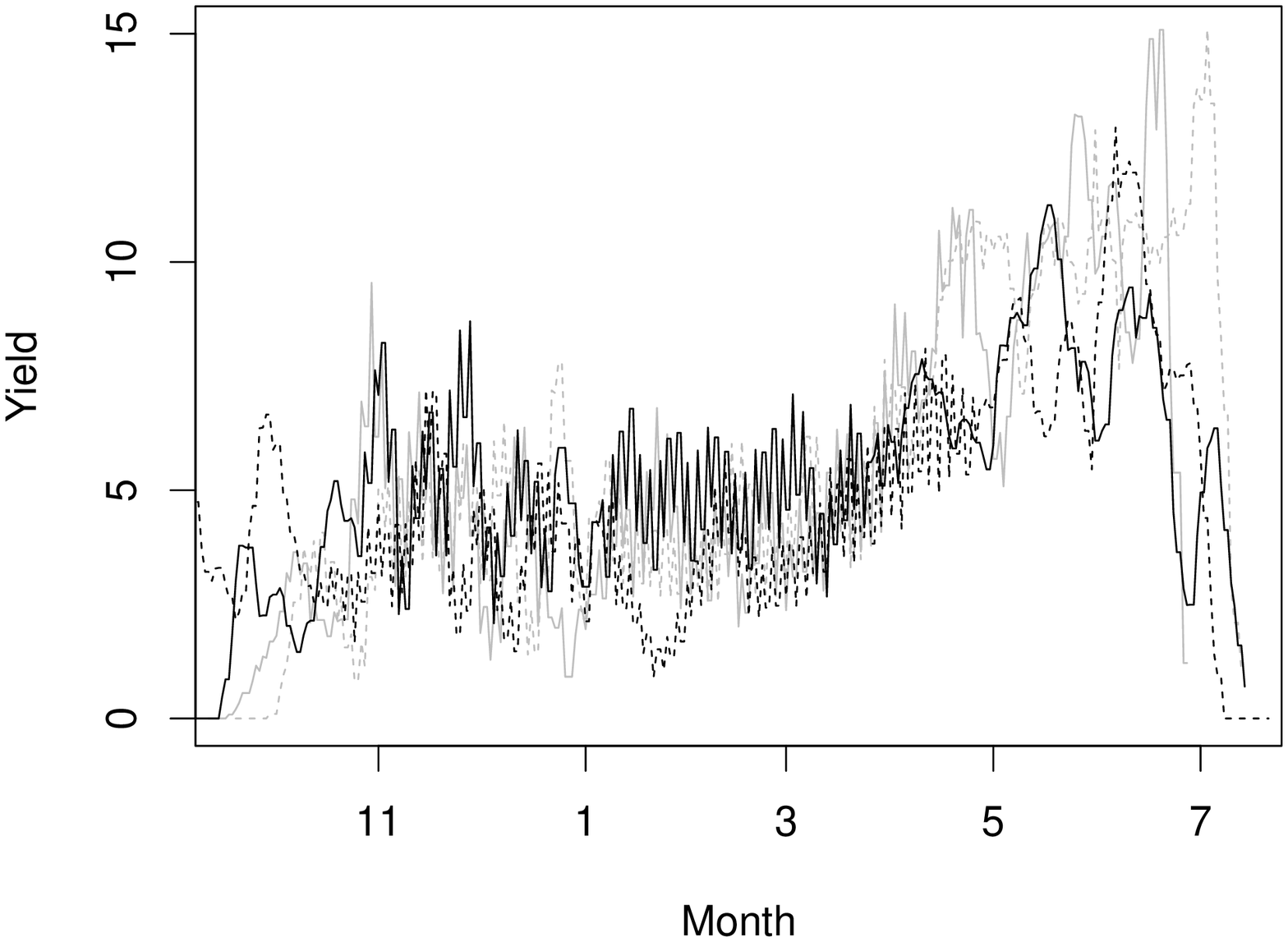}
		\includegraphics[width=0.48\hsize]{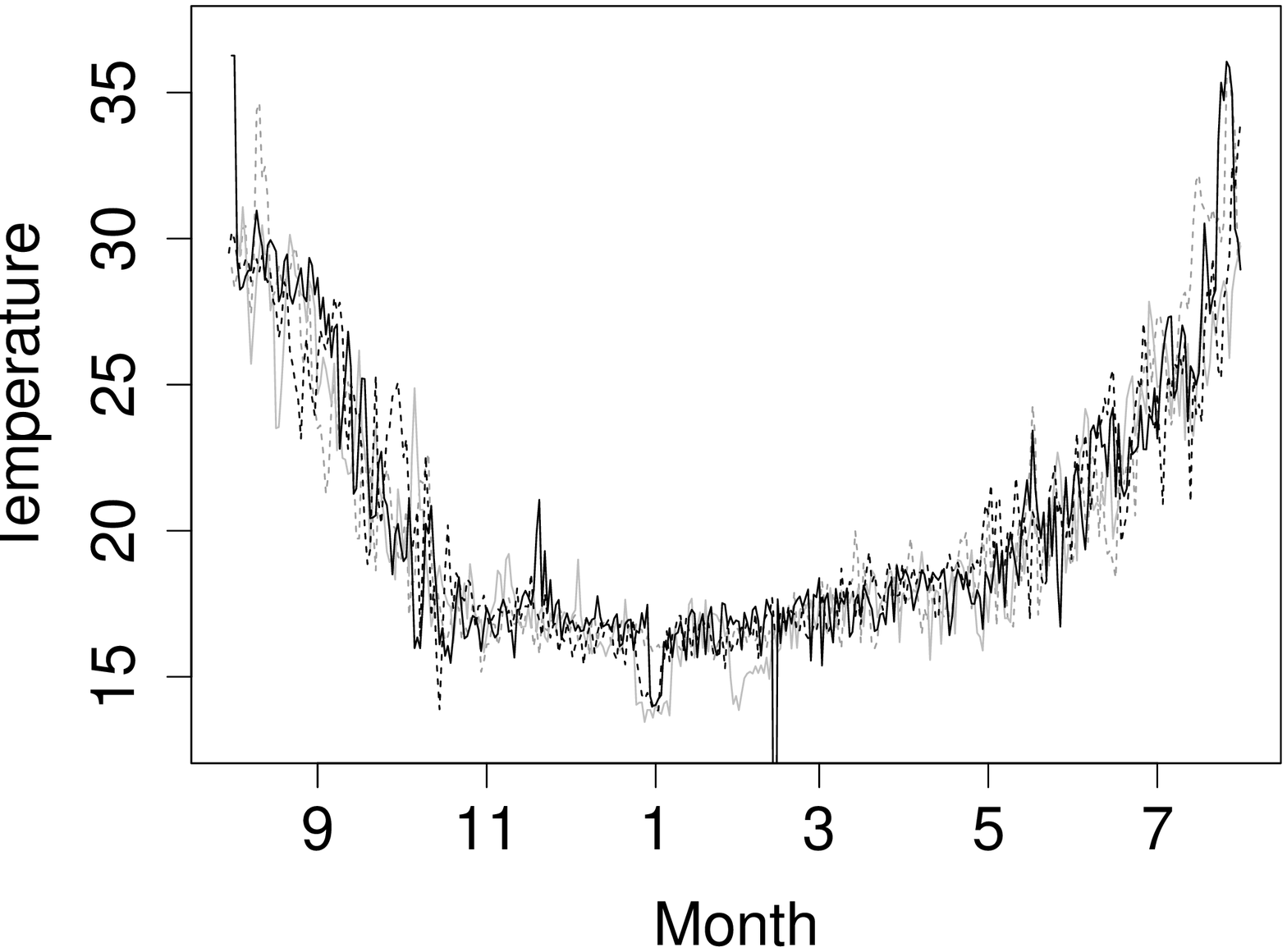}
	\end{center}
	\caption{Left: amount of crop yield for 4 terms, where the moving averages for 7 days are shown. Right: daily temperatures for 4 terms, where the range of the horizontal axis is from August 1st to July 31st. } 
	\label{fig:tomato_data}
\end{figure}

Each plant of the multi-stage tomatoes grows for about one year from August to next July, and are harvested almost every day from October to next July. 
In this study, we obtain daily yield data of one breed observed from October 2017 to July 2021, which is shown in Figure \ref{fig:tomato_data} left.  
The observed data for the daily yield vary greatly because the yield is 0 when the farm is on a holiday, which makes the analysis too difficult.  
Therefore, we use moving averages of the yield up to seven days from the harvest date as data.  
Environmental factors such as temperature and solar radiation are repeatedly measured by measuring equipment installed inside and outside the greenhouse, here we use the temperature inside the greenhouse as data for the environmental factor.
Data on environmental factors are observed every minute, here we treated daily averages as data.  
The daily averaged temperature from August, 2017 to July, 2021 is shown in Figure \ref{fig:tomato_data} right.  

It is considered by farmers that the growth of the tomato fruits is influenced by environmental factors during 60-day period before maturing the tomatoes.  
On the other hand, the period that relates to the yield is considered to vary with season; that is, the period that the temperature relates to the yield in summer differ from that in winter.  
This fact has not yet been assessed quantitatively.  
Therefore, we try to investigate how many days before maturing the tomatoes the temperature relates to the crop yield according to the seasons.  

First, we transformed the temperature data for 80 days before each day that the crop yield is observed into functional data by $B$-spline basis expansions.
Then we constructed the VCFLM, treating the daily yield of tomatoes as a response, the temperature corresponding to 80 days before the maturing day as a functional predictor, and the calendar time (From Jan. 1st to Dec. 31st) as an exogenous variable.  
A set of a yield of certain day and 80-day temperature before the day corresponds to an individual, and the sample size is $n=1172$.  
%Readers may think that the analysis of this dataset can be applied by the function-on-function regression model with sparsely observed data discussed in \cite{YaMuWa2005b}.
%In our case, however, the number of time points for the response is one for individual, and is not included in the function-on-function regression model.  
%If we have the crop yield data for decades of years, we may be able to apply the function-on-function regression models by treating the yearly crop yield data as individuals, but the observed period of the dataset is only three years.  
%For such dataset, our VCFAM is applicable.  
The VCFLM is estimated by the penalized likelihood method with the nested group bridge penalty.  and then the tuning parameters $\kappa$ and $\lambda$ in (\ref{eq:NGB}) by BIC.  
Here we fixed the other tuning parameters as $m^{(1)} = 15$, $m^{(2)} = 10,$ and $\gamma = 0.5.$ 

\begin{figure}[t]
	\begin{center}
		\includegraphics[width=0.7\hsize]{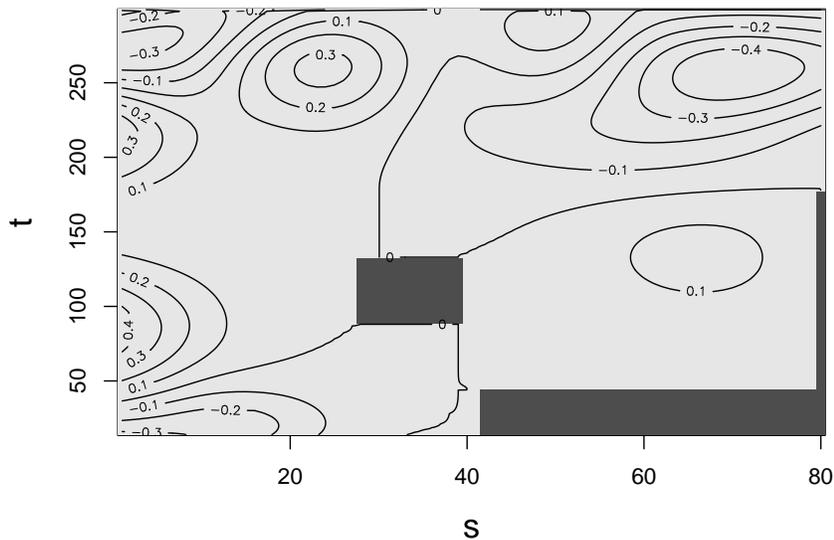}
	\end{center}
	\caption{Estimated coefficient surface $\beta(s,t)$ for the crop yield data.  The horizontal axis indicates the day before cultivation and the vertical axis indicates the calendar day, where the bottom and top correspond to Oct. 1st and next Jul. 20th, respectively. Dark gray represents areas where the coefficient is estimated to be 0.} 
	\label{fig:tomat_beta}
\end{figure}

Figure \ref{fig:tomat_beta} shows estimated coefficient surface $\beta(s,t).$
This figure indicates that the coefficients are shrunken toward exactly zero around the bottom right of the figure, due to the effect of the proposed method.  
From this result we can find that the temperature 40 days before cultivation relates to the crop yields in autumn, whereas that more than 40 days before cultivation does not.  
On the other hand, in mid-winter, spring and summer, the temperature 80 days before cultivation still relates to the crop yields.  
Particularly in spring and summer, when the crop yields increase significantly, the higher the temperature around 20 days before harvest, the higher the crop yield.
Furthermore, the lower the temperature around 70 days before harvest, the higher the yield.
In other word, higher crop yield in spring and summer may come from the temperature difference between around 20 days and 70 days before cultivation.  

In winter, however, temperatures 30 to 40 days before harvest do not relate to the yield, while earlier temperatures again relate to.  
This means that it fails to obtain a strictly truncated model.
The data used in this analysis indeed may have such relationship, but the estimation method needs to be improved if we want to obtain the strict truncated model.  

%The proposed method shows how many days ago the environmental factor relates to the crop yield by season.
%

%%%%%%%%%%%%%%%%%%%%%%%%%%%%%%%%%%%%%%%%%%%%%%%%%%%%%%%%%%%%%%%%%%%%%%%%%%%%%%%%%%
\section{Concluding remarks}
We have proposed a method for the truncated estimation procedure for varying-coefficient functional linear models that expresses up to what point in time the predictor relates to the response for any values of exogenous variable.  
To obtain the truncated model, we apply the basis expansion approach and the nested group bridge regularization to estimate the coefficient parameters given as functions, and then shrink a part of coefficient function toward exactly zero.  
Through simulation studies we showed the effectiveness of the proposed method.  
We then applied the proposed method to the analysis of crop yield data for multi-stage tomatoes, treating the seasonal time as an exogenous variable.  
We investigated up to what point the environmental factor relates to the crop yield at different season.  

In real data analysis in Section 6, we used only the temperature as a predictor.  
In practice, however, it is not realistic to assume that there is only one environmental factor that relates to the crop yield, and it is natural to assume there are multiple predictors.  
An extension to the model with multiple predictors is one of the future works to be addressed.  
Although the estimation strategy for it may be straightforward, the identifiability of the coefficient estimates need to be investigated.  
In addition, some asymptotic properties such as consistency should be addressed as future works.  

As indicated in the end of the real data analysis, we found that it may fail to construct a strict truncated model, that is, the coefficient surface may be estimated to be zero only at the closed domain.
To solve this problem, other methods such as estimation with other penalties may be applied.  

%%%%%%%%%%%%%%%%%%%%%%%%%%%%%%%%%%%%%%%%%%%%%%%%%%%%%%%%%%%%%%%%%%%%%%%%%%%%%%%%%%
\subsection*{Appendix: Proof of Proposition 1}

Let
\begin{align*}
    \tilde\ell_{\kappa,\tau}(\bm\eta) = 
    	- n\sum_{l=1}^{m^{(2)}}\sum_{k=1}^{m^{(1)}} c_{kl}^{1/\gamma}\eta_{kl}^{1-1/\gamma} \|\tilde{\bm b}_{A_{k},l}\|_1
		- n\tau \sum_{l=1}^{m^{(2)}}\sum_{k=1}^{m^{(1)}}\eta_{kl}
\end{align*}
a function consisting of only terms that depend on $\bm\eta$ in the penalized log-likelihood function (\ref{eq:NGB2}).  
Then the maximizer of $\tilde\ell_{\kappa,\tau}(\bm\eta)$ must have zero gradient, that is, 
\begin{align*}
    \frac{\partial\tilde\ell_{\kappa,\tau}(\bm\eta)}{ \partial\eta_{kl}} =  
    -n\left(1 - \frac{1}{\gamma}\right)\eta_{kl}^{-1/\gamma}c_{kl}^{1/\gamma}\|\bm b_{A_k,l}\|_1 -n\tau = 0.
\end{align*}
Then we have
\begin{align*}
    \eta_{kl} = \tau^{-\gamma}\left(\frac{1}{\gamma}-1\right)^{\gamma}c_{kl}\|\bm b_{A_k,l}\|_1^\gamma.
\end{align*}
Substituting it to $\tilde\ell_{\kappa,\tau}(\bm\eta)$, we have
\begin{align*}
    \tilde\ell_{\kappa,\tau}(\bm\eta) = - n\lambda \sum_{l=1}^{m^{(2)}}\sum_{k=1}^{m^{(1)}} c_{kl} \|\tilde{\bm b}_{A_{k},l}\|^\gamma,
\end{align*}
which coincides the third term of (\ref{eq:NGB}).  

\subsection*{Acknowledgment}
We appreciate the Higashibaba farm for providing the data for cultivating tomatoes.  
This work was supported by JSPS KAKENHI Grant Number 19K11858.

%\bibliography{C:/w32tex/share/texmf-dist/bibtex/bib/biblatex}

\end{document}